\documentclass[11pt]{amsart}
\usepackage{amsmath,amssymb,eucal}
\usepackage[all]{xy}
\usepackage{hyperref}
\begin{document}


\newcommand{\ie}{\textit{i}.\textit{e}.\,}
\newcommand{\eg}{\textit{e}.\textit{g}.\,}
\newcommand{\cf}{{\textit{cf}.\,}}

\newcommand{\half}{{\textstyle{\frac{1}{2}}}}
\newcommand{\A}{{\mathcal A}}
\newcommand{\B}{{\mathbb B}}
\newcommand{\C}{{\mathbb C}}
\newcommand{\Cs}{{C^*}}
\newcommand{\Q}{{\mathbb H}}
\newcommand{\R}{{\mathbb R}}   
\newcommand{\M}{{\mathbb M}}
\newcommand{\D}{{\mathbb D}}
\newcommand{\Mo}{{\overline{\mathbb M}}}  
\newcommand{\mc}{{{\mathbb{U}^c}}}
\newcommand{\bP}{{\mathbb P}}
\newcommand{\mo}{{\bar{M}}} 
\newcommand{\Gl}{{\rm Gl}}
\newcommand{\PGl}{{\rm PGl}}
\newcommand{\Sl}{{\rm Sl}}
\newcommand{\T}{{\mathbb T}}
\newcommand{\mU}{{\rm{U}(2)}}
\newcommand{\um}{{\mathfrak u}}
\newcommand{\bone}{{\bf 1}}
\newcommand{\SU}{{\rm SU}}
\newcommand{\SO}{{\rm SO}}
\newcommand{\bu}{{\bar{u}}}
\newcommand{\bv}{{\bar{v}}}
\newcommand{\cv}{{\bf v}}
\newcommand{\be}{{\bf e}}
\newcommand{\bi}{{\bf i}}
\newcommand{\bj}{{\bf j}}
\newcommand{\bk}{{\bf k}}
\newcommand{\bLa}{{\bar{\lambda}_\infty}}
\newcommand{\Z}{{\mathbb Z}}
\newcommand{\ve}{{\boldsymbol{\varepsilon}}}
\newcommand{\BW}{{\rm BW}}

\newcommand{\cs}{{C^*}}
\newcommand{\Maps}{{\rm Maps}}
\newcommand{\Hom}{{\rm Hom}}
\newcommand{\Pic}{{\rm Pic}}
\newcommand{\spiii}{{\rm Spin \: III}}
\newcommand{\III}{{\bf (III)}}
\newcommand{\IV}{{\bf (IV)}}
\newcommand{\Spin}{{\rm Spin}}
\newcommand{\bb}{{\boldsymbol{\beta}}}
\newcommand{\MC}{{\rm MC}}
\newcommand{\PGlH}{{{\rm PGl}_{\mathbb C}({\mathbb H)}}}
\newcommand{\PR}{{{\mathbb RP}^3}}
\newcommand{\U}{{\mathbb U}}
\newcommand{\bX}{{\bf s}}
\newcommand{\BV}{{V}}
\newcommand{\X}{{\bf X}}
\newcommand{\tmf}{{\rm tmf}}
\newcommand{\coker}{{{\rm coker} \;}}
\newcommand{\ab}{{\rm ab}}
\newcommand{\fH}{{\sf H}}

\parindent=0pt
\parskip=6pt

\title{At the boundary of Minkowski space}

\author[Jack Morava]{Jack Morava}

\address{Department of Mathematics, The Johns Hopkins University,
Baltimore, Maryland 21218}

\date{May 2022}

\begin{abstract}{The Cayley transform compactifies Minkowski space $\M$, realized as self-adjoint $2\times2$ complex matrices following Penrose, as the unitary group $\U(2)$. Its complement is a compactification of a copy of a light-cone as it is usually drawn, constructed by adjoining a bubble or $\C P_1$ of unitary matrices with eigenvalue $\pm 1$ at the ends of a lightcone at infinity.\medskip

\noindent The Brauer-Wall group of $\U(2)$ (\ie of fields of certain kinds of graded $\Cs$-algebras, up to projective equivalence) is $\Z_2 \times \Z$, defining an interesting class of nontrivial examples of Araki-Haag-Kastler backgrounds for quantum field theories on compactified Minkowski space. The second part of this paper extends such models to link presentations of more general spin four-manifolds.}\end{abstract}

\maketitle \bigskip

{\sc Part I : The Weyl/Cayley transform} \bigskip

This work began as an exercise in linear algebra, \ie to interpret stereographic projection 
\[
\M \ni \X \mapsto {\sf C}(X) := \frac{\X - i{\bf 1}}{\X + i{\bf 1}} \in \U(2) \cong \T \times_{\pm 1} \SU(2) \cong \Spin^c(3)
\]
(regarded as defined on the Penrose-Minkowski space of self-adjoint  $2 \times 2$ Hermitian matrices 
\[
\X := \left[\begin{array}{cc}
     x_0 + x_1 & x_2 - ix_3 \\
     x_2 + ix_3 & x_0 - x_1 
\end{array}\right] 
\]
with $x_* \in \R^{1,3}$) as a compactification. It was precipitated by David Mumford's recent review of current cosmological literature, in particular by his beautiful image \cite{24}(Fig 1) of our past light-cone.

In \S 1 we show that this Cayley compactifiction has a stratification 
\[
\U(2) \;  \cong \; \M \cup \M_\infty \cup \B
\]
in which $\M_\infty$ is a `light-cone at infinity', and $\B \cong \C P_1$ is a two-sphere of unitary matrices with eigenvalues $\pm 1$. The Cayley compactification of $\M$ maps to Penrose's, with the point at infinity on the light-cone at infinity blown up as a two-sphere $S^2 = \C P^1$, providing a plausible keystone or linchpin \cite{26} for constructions involving the Bondi-Metzner-Sachs group \cite{22} of classical general relativity.

Section 3 discusses fields of $C^*$-algebras over this stratification as a homotopy-theoretic setting for algebraic quantum field theory. The second part of this paper goes on to argue that both the geometric categories of three-manifolds and the algebraic categories of Hilbert space operators have homological dimensions roughly three, and pair in ways evoking a  duality between differential topology and quantum physics.\bigskip

{\bf \S 1} Recollections and calculations \bigskip

{\bf 1.1} Let $\Sl_2(\C) \subset M_2(\C)^\times$ be the subgroup of $2 \times 2$ complex matrices $T$ with determinant one; note that the map $T \mapsto T^*$ which sends a matrix to its conjugate transpose or adjoint is an antihomomorphism, and that the determinant of the conjugate transpose of a matrix is the complex conjugate of the determinant of the original matrix. Then $\SU(2) \subset \Sl_2(\C)$ is the maximal compact subgroup, composed of matrices of the form
\[
T = \left[\begin{array}{cc}
     u & v \\
     -\bv & \bu
\end{array}\right] 
\]
with $u = u_0 + i u_1,\; v = v_0 + i v_1 \in \C$ such that $\det T = |u|^2 + |v|^2 = 1$ (\ie unit length elements of the quaternions $\Q = \C \oplus \C\bj$), and let $\U(2)$ be the group of invertible $2 \times 2$ complex unitary matrices $U$ (such that $U^* = U^{-1}$); its Lie algebra $\um$ consists of {\bf anti}Hermitian complex $2 \times 2$ matrices. The exponential map of a connected compact Lie group is surjective, and any element of $U \in \U(2)$ can be expressed uniquely \cite{3}(Ch 9) as
\[
U = 
\left[\begin{array}{cc}
     u & v \\
     - \lambda \bv & \lambda \bu
\end{array}\right]
\]
with
\[
|u|^2 + |v|^2 = 1, \; \det U  = \lambda = e^{i\alpha} \in \T, \; \alpha \in [-\pi,+\pi] \;,
\]
defining a homeomorphism of $\U(2)$ with $S^1 \times S^3$. However, the group extension
\[
\xymatrix{
1 \ar[r] & \SU(2) \ar[r] & \SU(2) \times_{\pm 1} \T \cong \U(2) \cong \Spin^c(3) \ar[r]^-\det & \T \ar[r] & 1 }
\]
is nontrivial.

 \bigskip

{\bf 1.2} Let
\[
\M := \{ M \in M_2(\C) \:|\: M = M^* \}
\]
denote the real vector space of $2 \times 2$ complex Hermitian (self-adjoint) matrices
\[
M =
\left[\begin{array}{cc}
     M_{11} & M_{12} \\
     M_{21} & M_{22} 
\end{array}\right] =
\left[\begin{array}{cc}
     \mo_{11} & \mo_{21} \\
     \mo_{12} & \mo_{22} 
\end{array}\right]  = 
\left[\begin{array}{cc}
     \rho_+ & w \\
     \bar{w} & \rho_- 
\end{array}\right] =
\left[\begin{array}{cc}
     x_0 + x_1 & x_2 - ix_3 \\
     x_2 + ix_3 & x_0 - x_1 
\end{array}\right]
\]
(with $w \in \C$, and $\rho_\pm,x_i \in \R$). Penrose coordinates $\R \times \R^3 = \R^{1,3} \to \M$ identify
\[
\det M := q(M)=  x_0^2 - (x_1^2 + x_2^2 + x_3^2) \in \R 
\]
with the Lorentz-Einstein pseudometric of Minkowski space. 

If $M \in \M$ then its eigenvalues are real, so $i\bone \pm M$ is invertible. Let
\[
\sigma : \M \ni M \mapsto \frac{M - i\bone}{M + i\bone} = \sigma(M) \in \mU
\]
denote the Cayley transform: essentially, $-i$ times Riemannian stereographic projection. This clearly satisfies $\sigma(M)^* = \sigma(M)^{-1}$, and because
\[
\bone - \sigma(M) = \bone -  \frac{M - i\bone}{M + i\bone} = \frac{2i\bone}
{M + i\bone} \;,
\]
is invertible, a matrix in the image of $\sigma$ cannot have 1 as an eigenvalue, so $\sigma$ has a well-defined inverse
\[
\sigma^{-1}(U) \; := \; i \frac{\bone + U}{\bone - U} \in \M
\]
on that image, guaranteeing that $\sigma$ is an embedding.

The complement $\Mo_\infty = \U(2) - \sigma(\M) \cong S^3/S^0$ consists of unitary matrices which {\bf do} have 1 as an eigenvalue; in particular, they can be written as $\exp(iZ)$ with $Z$ self-adjoint and zero as an eigenvalue. \bigskip

{\bf 1.3} If $ U \in \Mo_\infty$ then 
\[
\det (U - \bone) = \det
\left[\begin{array}{cc}
            u - 1 &  v \\
     -\lambda \bv & \lambda \bu -1
\end{array}\right] = 1 - (u + \lambda \bu) + \lambda = 0 \;,
\]
so Trace $U = u + \lambda \bu = 1 + \lambda = 1 + \det U$. For example, $\lambda = 1$ implies $U = \bone$, but if $\lambda = - 1$ then
\[
U =
\left[\begin{array}{cc}
            u  &  v \\
           \bv & -\bu
\end{array}\right]
\]
has trace zero, so $u = u_0$ is real. There is thus a `bubble', a two-sphere $\B \subset \U(2)$ 
\[
u_0^2 + v_0^2 + v_1^2 = 1 \;,
\]
of such matrices. 

{\bf 1.4} The light-cone is the subset\begin{footnote}{We regard $\R \ni 0, \; \C \ni 0$ as basepointed spaces, with one-point compactifications $\R_+ = \bP^1(\R) \cong S^1, \; \C_+ = \bP^1(\C) \cong S^2$.}\end{footnote} 
\[
\M_0 := \{M \in \M \:|\: \det M = 0 \} \cong \R \ltimes \C_+ = (\R \times \C_+)/
(0 \times \C_+)
\]
of Minkowski space. It can be parametrized by stereographic projection 
\[ (x_0,z) \mapsto x_0(1,\bX(z)) = M_0(x_0,z)
\]
where
\[
\C_+ \ni z \mapsto \bX(z) := (1 + |z|^2)^{-1}(|z|^2 - 1,2z) \in \R^3 \cong
\R \times \C \;,
\]
\ie
\[
(x_0,z) \mapsto M_0(x_0,z) = k
\left[\begin{array}{cc}
     |z| & u \\
     \bu  & |z|^{-1}
\end{array}\right] \in \M_0 
\]
with $u = |z|^{-1}z$ and $k = 2(|z| + |z|^{-1})^{-1}x_0$. \bigskip

{\it {\bf Claim} The composition $\sigma^\perp := - \sigma \circ M_0$,
\[
\sigma^\perp : \R \ltimes \C \ni (x_0,z) \mapsto  \frac{\bone + iM_0}{\bone 
- iM_0} \in \Mo_\infty \subset \mU
\]
is an embedding, with the} light-cone $\M_\infty$ at infinity {\it as its image, disjoint from $\sigma(\M)$.} \bigskip

In particular, $\sigma^\perp(0,z) = \bone$. The map is well-defined, for
\[
\det (\bone - iM_0(x_0,z)) = \det
\left[\begin{array}{cc}
     1 - ik|z| & - iku \\
     - ik\bu & 1 - ik|z|^{-1} 
\end{array}\right] =  (1 - ik|z|)(1 - i k|z|^{-1}) + k^2
\]
\[
= 1 - ik(|z| + |z|^{-1}) = 1 - 2ix_0 \neq 0 \;.
\]
This implies that 
\[
\bone - \sigma M_0 = 2(\bone - i M_0)^{-1}
\]
is invertible, and hence that $\sigma^\perp$ is an embedding since
\[
M_0 = i \frac {\bone + \sigma M_0}{\bone - \sigma M_0} \;.
\]
The image of $\sigma^\perp$ is disjoint from $\sigma(\M)$, because
\[
\det (\bone + \sigma M_0) = \det \frac{2iM_0}{1 - iM_0} = 0
\]
implies $-\sigma M_0$ has 1 as an eigenvalue. \bigskip

{\bf 1.5} Calculation now shows that
\[
(1 - 2ix_0) \sigma^\perp (x_0,z) =
\left[\begin{array}{cc}
     1 + ik|z| & iku \\
     ik\bu & 1 + ik|z|^{-1}
\end{array}\right] \cdot
\left[\begin{array}{cc}
     1 - ik |z|^{-1} & iku \\
     ik\bu & 1 - ik|z|
\end{array}\right] =
\]
\[
\left[\begin{array}{cc}
     1 + ik(|z| - |z|^{-1}) & 2iku \\
     2ik\bu  & 1 + ik(|z|^{-1} - |z|)
\end{array}\right] = \bone + 2ix_0 \BV(z) \;,
\]
where
\[
\BV(z) = (|z| + |z|^{-1})^{-1}
\left[\begin{array}{cc}
     |z| - |z|^{-1} & 2u \\
     2\bu & |z|^{-1} - |z|
\end{array}\right]
\]
is Hermitian, satisfying $\BV^2 = \bone$ and Trace $\BV = 0$. If 
$z = re^{i\theta}$, then
\[
\BV(re^{i\theta}) =
(r^2 + 1)^{-1} \left[\begin{array}{cc}
                       r^2 - 1 & 2r e^{i\theta} \\
                      2r e^{-i\theta} & 1 - r^2
               \end{array}\right] \;.
\]

Evidently $P = \half (\bone + \BV)$ is an element of the space $\D$ of projections with Trace $P = 1$ and $\be = (z,1) \in \C^2$ as eigenvector. We have  
\[
\sigma^\perp(x_0,z) = \frac{\bone + 2ix_0 \BV(z)}{1 - 2i x_0} = \bone + \frac{4ix_0}{1 - 2ix_0} P \;,
\]
so Trace $\sigma^\perp = (1 - 2ix_0)^{-1} = 1 + \det \sigma^\perp$, \ie
\[
\det \sigma^\perp(x_0,z) = \frac{1 + 2ix_0}{1 - 2ix_0} = e^{i\alpha(x_0)}\ \in \T 
\]
with
\[
x_0 = -\half \tan \half \alpha \;, \alpha(\pm \infty) = \pm \pi \;.
\] 

If we write $- \beta$ for 
$\frac{4ix_0}{1 - 2ix_0} = e^{-i\alpha} - 1$, then $\sigma^\perp(x_0,z) = \bone - \beta P$, so
\[
\log (\bone - \beta P) = - \sum_{n \geq 1} \frac{(\beta P)^n}{n} = \log (1 - \beta) \cdot P = - i \alpha P 
\]
and hence 
\[
\sigma^\perp(x_0,z) = \exp(- i\alpha P) \;.
\]

This identifies the space $\D$ of projections with the bubble of unitary matrices with eigenvalues $\pm 1$.

Let
\[ 
\ve :=  \left[\begin{array}{cc}
                     1 & 0 \\
                     0 & -1
          \end{array}\right] \;,
\]
then $\BV(z) \to \mp \ve$ as $z \to 0$ resp $\infty$. Similarly, as $x_0 \to 0, \; \sigma^\perp(x_0,z) \to \bone$, while
\[
\sigma^\perp(x_0,z) \to - V(z) \in \B = \Mo_\infty - \M_\infty
\]
as $x_0 \to \pm \infty$, so $\sigma^\perp(x_0,z) \to \ve$ as $(x_0,z) \to (\infty,\infty)$.\bigskip

{\bf Remark} If $\B = \left[\begin{array}{cc}
                                    b_+ & w \\
                                \bar{w} & b_-
                             \end{array}\right] \in M_2(\C)$ is Hermitian,
with determinant zero and trace one, then it is a projection. Setting $r = (1 - b_+)^{-1}|w|$ identifies it with $\D$. \bigskip

{\bf 1.6} It follows that $\sigma^\perp$ extends to a homeomorphism 
\[
\bar{\sigma}^\perp : \R_+ \ltimes \C_+ \cong (\R_+ \times \C_+)/
(0 \times \C_+) \to \Mo_\infty \;.
\]
Note that the domain of this map can be expressed as 
\[
\R_+ \wedge (* \sqcup \C_+) \cong \Sigma(S^0 \vee S^2) \;,
\]
where $\Sigma$ denotes the reduced suspension used in homotopy theory. \bigskip

{\bf Corollary}\begin{footnote}{It is not clear to me how well this is understood in the physics community; \cf. 
\cite{12}(\S 5.1). I learned of \cite{16} only after posting an earlier version of this paper.}\end{footnote} {\it The obvious inclusion induces an isomorphism $H^*(\U(2),\Z) \cong H^*(\Mo_\infty,\Z)$ in degrees below four; moreover,
\[
\Mo_\infty - \M_\infty \cong S^2, \; \U/\Mo_\infty \cong S^4 \;,
\]
while $H^*(\U(2)/\B,\Z) = \Z$ when $* = 3,4$ and is zero otherwise.}\bigskip

{\bf An exercise}, with most grateful thanks to David Mumford:

As  $t \to \infty$, a light ray $x_*(t) = (0,{\bf x}) + t(1,{\bf v}) \in \R \times \R^3$ (with $|{\bf v}| = 1$) approaches 
\[
\left[\begin{array}{cc} 
u & v \\
-\lambda \bar{v} & \lambda \bar{u} 
\end{array}\right]
 =
\frac{1}{1 - i \omega}  
\left[\begin{array}{cc}
z & - \nu \\
 - \bar{\nu} & - \bar{z}  
 \end{array}\right] \in \U(2)
 \]
 as above, 
with $\omega := {\bf x \cdot v},\; \lambda = {\sf C}(\omega), \; z = v_1 + i\omega,\; \nu = v_2 + i v_3$, ending on the line 
\[
v = - \frac{v_2 + i v_3}{1 + iv_1}(1 + iu), \; |u|^2 + |v|^2 = 1.
\]
\bigskip

{\bf \S 2} Some group actions \bigskip

{\bf Definition}
\[
\Sl_2(\C) \times \M \ni T,M \mapsto T(M) := TMT^* \in \M
\]
defines a group action: for
\[
(T(M))^* = (TMT^*)^* = TM^*T^* = T(M) \;,
\]
while
\[
S(T(M) = S(TMT^*)S^* = (ST)M(ST)^* = (ST)(M) \;.
\]
Moreover,
\[
\det(T(M)) = \det(TMT^*) = \det T \cdot \det M \cdot \det T^* = \det M \;.
\]

{\bf Corollary} {\it $\Sl_2(\C)$ is the double cover of the identity component  of the (Lorentz) group of isometries of $(\M,q)$.} \bigskip

The action of the subgroup $\SU(2)$ on $\M$ preserves the decomposition of $\M$ into (Time) $\times$ (Space), factoring through the action of the rotation group $\SU(2) \to \SO(3)$ on the second term. Moreover, the conjugation action of $\SU(2)$ on $M_2(\C)$ defined by the composition
\[
\SU(2) \to \Sl_2(\C) \to {\rm PGl}_2(\C)
\]
preserves the matrix algebra structure.

By the remarks in the previous section, $\sigma$ is equivariant with respect to the action of $\SU(2)$ on $\U(2)$ by conjugation.

The action of $\SU(2)$ on $\B = \Mo_\infty - \M_\infty$, regarded as the space of projections in $M_2(\C)$ with determinant zero and trace one, can be identified with its action via $\PGl_2(\C)$ on the space of projections with eigenvector $\be = [z:1]\in \bP_1(\C)$, defining a Hopf bundle at time-like infinity. This is reminiscent of (the other kind of Hopf) bifurcation.\bigskip

{\bf \S 3} A sandbox for entanglement\bigskip

{\bf 3.1} The Brauer-Wall/Maycock group 
\[
0 \to H^3(Z,\Z) \to ({\rm BW} \cong \MC)(Z) \to H^1(Z,\Z_2) \to 0
\]
(with composition $(b,s) + (b',s') := (b + \beta(s \cdot s') + b', s + s')$, \cite{39}(Prop 2.5), $\beta$ being the mod two Bockstein; represented by a truncation of the loopspace $\Omega^\infty k{\mathbb O}$) classifies Morita equivalence classes of fields of graded continuous trace class $\cs$ algebras over a $CW$-space $Z$. 

Contractibility of the group of invertible Hilbert space operators implies that bundles $H^1(Z,{\rm PGl}_\C({\mathbb H}))$ \cite{5, 8, 9, 23, 40}  of projective Hilbert spaces over $Z$  -- equivalently, locally coherent fields of quantum mechanical state spaces -- are classified by elements of
\[
H^3(Z,\Z) \cong H^2(Z,B\Z \simeq \T) \cong H^1(Z,B\T \simeq {\rm Gl}_\C({\mathbb H})/\C^\times) \;. 
\] 
Small $H$-spaces $H(V,1) \rtimes_q H(\Z_2,3)$ generalizing $\MC$ can be associated naturally to symmetric bilinear forms $q : V \times V \to \Z_2$ in characteristic two; \cf \S6.

For the purposes of this note, a Haag-Kastler background $[\A]$ on a connected locally compact space $Z$ will be the projective equivalence class of a bundle of complex Hilbert spaces trivialized at infinity on its one-point compactification $Z_+$, as a toy model for quantum mechanics. Compactly supported cohomology groups $H^*_c(Z) := H^*(Z_+,+)$ (\ie defined by the one-point compactifications of the components of $Z$) are useful in this context; the resulting functors are natural with respect to proper, but not general, homotopy equivalence.

{\bf Proposition} {\it A connected oriented three-manifold $Y$ has a canonical Haag-Kastler background $[\A_Y]$
of $\cs$ algebras defined by its orientation or volume form $[\omega_Y] \in H^3_c(Y,\Z)$.} 

The light-cone $\M_0$, for example, is contractible, but its two ends imply a serious amount of compactly supported cohomology: 
\[
H^*_c(\M_0,\Z) \cong \Z \; {\rm if} \; *= 1, \; \cong \Z^2 \; {\rm if} * = 3 
\]
and is otherwise zero; and, similarly, by \S 1.4, for $\M_\infty$. A chiral structure on the light-cone \cite{28} is defined by a choice of the isomorphism in degree three; it is not clear to me that the two ends need necessarily to be glued by the identity map. Collapsing $\B$
\[
\Mo_\infty = \M_\infty \cup \B \to \M_{\infty +} 
\]
sends $H^3_c(\M_\infty) \cong \Z^2 \to \Z \cong H^3_c(\Mo_\infty)$. The decomposition $\M_\infty = \Mo_\infty - \B$, together with the long exact sequence 
\[
\xymatrix{
\dots \ar[r] & H^*_c(X-Z,\Z) \ar[r] & H^*_c(X,\Z) \ar[r] & H^*_c(Z,\Z) \ar[r] & \dots }
\]
for a closed subspace $Z \subset X$ then implies an exact sequence
\[
\xymatrix{
0 \ar[r] & {H^2_c(\B,\Z) \cong \Z} \ar@{.>}[r]  & {\BW(\M_{\infty +}) \cong \Z^2} \ar[r] & {\BW(\Mo_{\infty}) \cong \Z_2 \times \Z}  \ar@{.>}[r]  & {H^2_c(\B,\Z_2) \cong \Z_2} \ar[r] & 0 \;.}
\]

The restriction of $\A$ to $\M$ is trivial since $\BW(\M_+) = 0$, but an algebra bundle of class $[\A]$ over $\U(2)$ nevertheless defines at least a precursor for a Haag-Kastler structure: it provides a sheaf of $\Cs$-algebras and quantum-mechanical state spaces, though without any concerns \cite{38} about local causality. This is an issue of possible interest in questions of entanglement. \bigskip

{\bf Corollary} {\it There is a canonical nontrivial equivalence class $[\A]$ of bundles of $\Z_2$-graded $\Cs$-algebras over $\U(2)$, classified by 
\[
(-1,+1) \in \Z_2 \times \Z \cong H^1(\U(2),\Z_2) \times H^3(\U(2),\Z) \cong \BW(\U(2)) \;. 
\]
This bundle is supported on $\Mo_\infty$, in the sense that the restriction map
\[
\BW(\U(2)) \to \BW(\Mo_\infty)
\]
is an isomorphism.}

The final arrow in the exact sequence above similarly suggests that the spin part of the structure is supported on the bubble $\B$. The Bockstein homomorphisms for both spaces are trivial.\bigskip

{\bf 3.2} Some questions: This document is a working draft; it is intended to provide a framework for questions like the following:

$\; \bullet$ Is there an analytic construction for (a bundle of class) $[\A]$?\\
$\; \bullet$ Does the class $[\A]$ contain a smooth representative?\\
$\; \bullet$ Can the action of $\SU(2)$ on $\U(2)$ be extended to some algebra bundle representing $[\A]$?\\

[More precisely: can $[\A]$ be realized as the bundle of automorphisms of a field of (projective) Hilbert space representations of $\SU(2)$ over $\U(2)$? If so, could these be related to (projective) representations of $\Sl_2(\C)$?]

The Bondi-Metzner-Sachs group \cite{22} is a semi-direct product
\[
0 \to V \to {\rm BMS} \to \Sl_2(\C) \to 1 \;,
\]
where $V$ is a vector space of real-valued functions on $\C P_1$ with the induced $\PGl_2(\C)$ action; it is the symmetry group of a generic asymptotically-flat solution of the equations of general relativity. It is tempting to imagine $V$ as the group of smooth functions on $\B$, interpreted as conformal deformations of its metric.

$\; \bullet$ \cite{25} How is a principal bundle ${\rm PGl}^*(\A) \to \U(2)$ related to $\T \times_{\pm 1} S^3 \langle 3 \rangle$? \bigskip

{\sc Part II An ocean of three-manifolds}\bigskip

\begin{quotation}{\dots Nehwon is a giant bubble rising through the waters of eternity with continents, islands, and the great jewels that at night are the stars all orderly afloat on the bubble's inner surface \dots \medskip

\noindent F Leiber, {\it  Swords of Lankhmar}}\end{quotation}\bigskip

{\bf 4.1} Following N Strickland \cite{29}(\S 12-13) and GA Swarup \cite{30, 31}, the category $\III$, with compact connected closed base-pointed oriented three-manifolds $Y$ as objects, and with degree one maps as morphisms, maps fully faithfully by $Y \mapsto \pi_1Y$ to the category of groups $\pi$ endowed with the three-dimensional level\begin{footnote}
{with the three-form defined by the triple product \cite{33} on $H^*B\pi$ as something like a Cartan connection}\end{footnote} structure $H_3Y \to H_3B\pi_1(Y)$, and with homomorphisms of such oriented groups as morphisms. The three-sphere $S^3 = \SU(2)$ is a distinguished point of this generalized stack, as is $S^1 \times S^2$, but the generic example of a prime object under connected sum is an acyclic three-manifold with fundamental group satisfying three-dimensional Poincar\'e duality. There is also an archipelago of manifolds such as Lens spaces, which have finite fundamental groups.

For example, the collapse map $S^1 \times S^2 \to S^1 \wedge S^2 \cong S^3$ has degree one. It changes the Kervaire semicharacteristic mod two \cite{10}.\bigskip

{\bf 4.2} On another hand, the Morita equivalence classes $\MC(Y)$ define a sheaf of abelian groups on $\III$, and the Grothendieck category
\[
({\rm HK}) := \int_{Y \in \III} \MC(Y) 
\]
of compact three-manifolds, together with the $\cs$ algebra indexed by their orientations $[Y] \in H^3(Y,\Z)$, defines an interesting class of background geometries for Araki-Haag-Kastler models.  Bundle gerbes \cite{17} and Deligne cohomology provide smooth versions of these things, in terms of connections and curvature.\bigskip

An element $(b,s) \in \MC(Y)$ defines the class $s \in H^1(Y,\Z_2)$ of a spin or fermionic structure, together with a class $b \in H^3(Y, \Z)$ which could perhaps be called a boson or baryon number, but that may be misleading. From here on we'll restrict our attention to the cross-section ${\rm (HK)}_1$ of the category of Haag-Kastler models defined by normalizing at $b = 1$. \bigskip

{\bf 5.1} More generally, let us consider the category $\IV$ with pairs $(Y \cong \partial X \subset X)$ as objects, with $X$ a connected oriented smooth four-manifold bounded by $Y \in \III$, and smooth maps of pairs with boundary restrictions of degree one, as morphisms. Forgetting the spanning manifold defines a fibration 
\[
\partial : \int_{X \in \IV} \MC(X) \to \int_{Y \in \III} \MC(Y)
\]
of some kind of categories. 

If $X$ is simply-connected, the homology exact sequence of $(X,Y)$ reduces (using the universal coefficient theorem and Lefschetz duality as in Hatcher [\S 3.3]) to a free three-term resolution 
\[
0 \to H_2Y \to H_2X \to H_2X/Y \to H_1Y \to 0
\]
 of $H_1Y$ (coefficients are integral if unspecified), and thereby a contravariant class 
 \[
 {\mathcal Q}_{Y:X} \in {\rm Ext}^2_{\Z[\pi]} (\pi_\ab,\pi_\ab^\dagger)
 \]
 \cite{11}(\S 5.3.13f). Here $A^\dagger := \Hom(A,\Z)$ for finitely generated abelian groups, and $\pi = \pi_1Y$. The diagram 
\[
\xymatrix{
\Hom(H_2X/Y,\Z)  \ar[r] & \Hom(H_2X,\Z) \\
H^2X/Y \ar[u]_\cong \ar[d]^\cong \ar[r] & H^2X \ar[u]_\cong \ar[d]^\cong \\
H_2X \ar@{.>}[uur]^-{Q_{X/Y}} \ar[r] & H_2X/Y \ar[r] & H_1Y \ar[r] & 0 }
\]
identifies the unimodular intersection form $Q := Q_{X/Y}$ on $H_2X$ with that defined by the cup product on $H^2X/Y$, yielding a presentation 
\[
H_1Y = \pi_ \ab  \cong \coker Q \;,  H_2Y = \pi_\ab / {\rm tors} \cong \ker Q
\] 
of $H_*Y$ in terms of a quadratic form.\bigskip

{\bf 5.2} Link calculus \cite{11, 18, 21},\cite{27}(Ch 9 \S I) presents any $Y \in \III$ as the boundary $Y \cong Y_L$ of a {\bf simply-connected} four-dimensional handlebody $X_L$ defined by a framed oriented link 
\[ 
L = \bigcup_{\lambda \in \pi_0 L} \lambda \subset \R^3_+ \;,
\]
together with an identification of the intersection matrix of $X_L$ and the $\pi_0 L \times \pi_0 L$ linking matrix of $L$.
\bigskip

It is helpful to know that the Stiefel-Whitney map
\[ 
\Pic_{\otimes \R}(Z) \ni \xi \mapsto w_1(\xi) \in H^1(Z,\Z_2) 
\]
classifies real line bundles, while Chern's map
\[
\Pic_{\otimes \C}(Z) \ni \lambda \mapsto c_1(\lambda) \in H^2(Z,\Z)   
\]
classifies complex line bundles. In a link presentation, equivalence classes 
\[
\lambda \in H_2X_L \cong H^2(X/Y)_L \cong \Pic_{\otimes \C}(X/Y)_L \cong \Z[\pi_0L] := \Lambda \cong \Z^l
 \]
correspond to line bundles $\lambda$ over $X$ trivialized on $Y$, or to the surfaces $[\sigma^{-1}(0)] \sim \delta_\lambda \in H_2(X_L) $ defined by the Euler class of a generic section $\sigma$. 

With $\Z_2$ coefficients, and in cohomology $\fH$ for convenience, the exact sequence of \S3 becomes a symmetric biextension
\[
\xymatrix{
0 \ar[r] & \fH^1Y \ar[d]^\cong \ar[r]^-\delta & [\fH^2X/Y \sim \fH^2X] \ar[d]^\cong \ar[r] & \fH^2Y \ar[d]^\cong \ar[r] & 0 \\
0 \ar[r] & \Pic_{\otimes \R}(Y) \ar[r]  & [\Pic_{\otimes \C}(X/Y) \otimes_\Z \Z_2, q_{X/Y}] \ar[r] & \Pic_{\otimes \R}(Y)^\vee \ar[r] & 0 }
\]
of $\Z_2$ - vector spaces (with $\vee$ denoting vector space duality and $ q := Q \otimes \Z_2$). The left-hand monomorphism sends a real vector bundle $\xi$ on the boundary $Y$ to a complex line bundle $\delta_\xi$ on X; we may perhaps interpret it as bosonic $\C^\times$-gauge field on the interior created or supported by a fermionic field on the boundary:

A generic section $\sigma$ of a real line bundle over $Y$ defines a codimension one submanifold $\sigma^{-1}(0)$, whereas such a section of a complex line bundle over $X$ defines (mod two) a submanifold of codimension two \cite{14}(lemma 5.49) but these submanifolds are both surfaces, making it geometrically natural to think of a class in $H^2(X,\Z) \otimes \Z_2$ as extending  a class in $H^1(\partial X,\Z_2)$ when its associated complex field turns on.\bigskip 

{\bf 6.1} This leads to matters of  spin and statistics, which suggests a pullback
\[
\xymatrix{
({\rm GR}) \ar@{.>}[d] \ar@{.>}[r] & ({\rm HK}) \ar[d] \\
\IV_{\Spin^c} \ar[r] &  \III_\Spin } 
\]
of our fibered category. The geometry of link calculus on Spin and $\Spin^c$ manifolds is rich enough to support 
(renormalizeable \cite{1} quantum) variational problems of Seiberg-Witten, Higgs-Yamabe \cite{42} and Salam-Weinberg type; the latter model involves mysterious  $\T$-valued `mixing angles' \cite{41} which parametrize interactions between fermions and their gauge bosons. 

This may be commensurable with Penrose's memorable fancy \cite{32}, that at future infinity fermions decay into bosons, powering a new big bang. A cobordism $Y = \partial X$ can be regarded as a creation operator $X : \emptyset \to Y$ which thinks of the four-manifold $X$ as a bubble blown by its boundary $Y$, a solution extremizing a functional on a moduli space of membranes spanning a given boundary.   

If $Y$ is $S^3$ then $X$ is a 4-ball with $X/Y = S^4$, and when $Y = S^1 \times S^2$ we have $X = S^1 \times B^3$, $X/Y \cong \U(2)$, which recovers Penrose's model. [In that case $X$ is not simply connected, but can be made so by allowing a codimension two singularity in $Y \sim S^1 \times S^2/\infty \times S^2$, \cf \S1.6.]\bigskip

{\bf 6.2} This marks a place for a discussion of spin links \cite{11}(\S5.7.11),\cite{18}(App. C),\cite{34} which we defer to a later draft. The following needs expansion and details:

In a link presentation $X_L$, a generic section of complex line bundle $\lambda$ defines the homology class $\delta_{L}$ of a (for example `weak neutral') de Rham current, normal to its vanishing locus $\sigma^{-1}(0)$: a Dirac delta-function supported by the link, a model for a thunderbolt\begin{footnote}{visible at the base of Figure 1 in \cite{24} if you look hard enough.}\end{footnote} or crack of doom in the big bang. 


Kirby and Taylor use the bilinear form $x,y \mapsto x^2y$ on $\Pic_{\otimes \R} Y$ to show that the $\xi$-twisted Rokhlin \cite{33}(Theorem VI) invariant  
\[
\nu(\xi^*Y) \; \equiv \;  \nu(Y) + 2 \bb(\star \xi) \; \; (mod \; 16)
\]
of a spin three-manifold is translated by a multiple of the EH Brown invariant \cite{4}(\S 4.2, 5.4), \cite{19}(\S 3.2, 4.11) of the surface Poincar\'e dual to $\xi$. Hopkins and Singer \cite{7}(App. E) study such refinements of the intersection matrix in terms of integral Wu classes; we hope to understand this better, in time. \bigskip \bigskip

\bibliographystyle{amsplain}

\end{document}